# Galactic Archaeology: Tracing the Milky Way's Formation and Evolution through Stellar Populations

Collazos Rozo J. Alfredo[1]

[1]*Peter the Great St. Petersburg Polytechnic University, Politechnicheskaya 29, 195251, St. Petersburg, Russia*
*PhD student at the Institute of Physics and Mechanics*
*Master's Degree at the Institute of Physics, Nanotechnology and Telecommunications*
*Physical Engineer of the Technological University of Pereira - Colombia*
*E-mail: jacollazos@utp.edu.co*

## ABSTRACT

Galactic archaeology represents a multidisciplinary approach aimed at unraveling the intricate history of the Milky Way galaxy through the study of its stellar populations. This abstract delves into the significance of galactic archaeology as a vital tool for understanding the formation and evolution of the Milky Way. By examining the distribution, kinematics, chemical compositions, and ages of stars within the galaxy, researchers gain insights into the dynamic processes that have shaped its structure over billions of years.

Stellar populations serve as invaluable relics of past epochs, preserving clues about the conditions prevalent during their formation. The utilization of spectroscopic and photometric data has enabled the classification and analysis of stars, allowing astronomers to discern distinct populations and unveil their origin stories. Through these studies, the emergence of the Milky Way's various components, such as the thin and thick disk, halo, and bulge, becomes discernible.

**Key words:** Galactic Archaeology - Milky Way - Stellar populations –Formation - Evolution.

## 1 INTRODUCTION

The Milky Way, our home galaxy, has long captivated the imagination of astronomers and scientists as a dynamic and evolving cosmic structure. Understanding its formation and evolution is not only an intellectual pursuit but also essential for comprehending the broader mechanisms governing galaxy evolution in the universe. One of the key methodologies that have enabled the unraveling of the Milky Way's history is the burgeoning field of galactic archaeology, which employs stellar populations as archaeological relics to reconstruct the galaxy's past. By analyzing the distribution, kinematics, chemical abundances, and ages of stars within the Milky Way, researchers have begun to reconstruct a detailed narrative of its formation and evolution. (Hawthorn & Gerhard 2016).

The concept of galactic archaeology draws inspiration from terrestrial archaeology, where the examination of artifacts and structures provides insights into the history and development of human civilizations. Similarly, in the realm of astronomy, the stars within the Milky Way can be seen as cosmic "fossils" that contain a wealth of information about the conditions and processes present during their formation. As these stars traverse various regions of the galaxy, they accumulate characteristics that reflect the galactic environment at the time of their birth (Renfrew & Bahn 2015).

Historically, the study of stellar populations and their roles in understanding the Milky Way's history dates back to the mid-20th century. The pioneering work of astronomers such as Jan Oort and Walter Baade laid the groundwork for investigating the dynamics and kinematics of stars in the Milky Way's halo and disk. As observational techniques advanced, more detailed information about stellar compositions and ages became accessible, enabling researchers to refine their understanding of the Milky Way's evolution (Sellwood & Binney 2002).

The utilization of spectroscopy and photometry has been instrumental in the classification and analysis of stars, revealing the existence of different populations with distinct properties. The ongoing advancements in astronomical instruments, coupled with extensive surveys, have yielded unprecedented amounts of data on stellar properties, facilitating the categorization and examination of stellar populations on a grand scale (Sozzetti & Bonavita 2018).

This paper aims to delve into the fundamental principles of galactic archaeology and its role in tracing the formation and evolution of the Milky Way through the study of stellar populations. By delving into the distribution, kinematics, chemical compositions, and ages of stars, researchers have gained insights into the galaxy's structural components and the processes that have shaped its present-day appearance. Through a comprehensive review of relevant literature and the exploration of key findings, this paper seeks to highlight the significance of galactic archaeology in elucidating the Milky Way's intricate history (Ting Y. S 2017).

# 1 STELLAR POPULATION CLASSIFICATION

Stellar population classification is a fundamental aspect of galactic archaeology, providing insights into the structure, history, and evolution of the Milky Way galaxy. Through the analysis of various stellar populations, astronomers gain a deeper understanding of the different components that make up our galaxy, each offering a unique glimpse into its past. This report delves into the classification of stellar populations, focusing on thin and thick disk stars, halo stars, bulge stars, and stellar streams and substructures (Bland Hawthorn & Gerhard 2016).

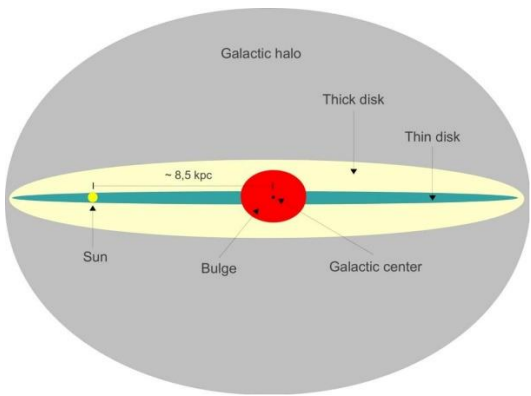

**Fig 1**. Edge on view of the Milky Way with several structures indicated (not to scale). The thick disk is shown in light yellow.

**Thin and Thick Disk Stars**: Thin and thick disk stars are two primary components of the Milky Way's disk structure. The thin disk comprises relatively younger stars that exhibit a well-defined plane with low vertical dispersion. These stars are typically metal-rich and are thought to have formed from a relatively calm process of gas accretion and star formation. The thick disk, on the other hand, consists of older stars that have a higher vertical dispersion. These stars are relatively metal-poor and are believed to have formed from a more turbulent process, possibly involving mergers or interactions. Studying the properties of thin and thick disk stars provides insights into the formation history of the disk and the conditions prevalent during their respective epochs (Bovy J 2012).

**Halo Stars:** Halo stars constitute the oldest component of the Milky Way, characterized by a spherical distribution around the galactic center. These stars have low metallicities and often exhibit distinct kinematics compared to disk stars. Halo stars are thought to have formed in the early universe, possibly through the accretion of smaller satellite galaxies or the early collapse of gas clouds. By analyzing the chemical abundances, kinematics, and ages of halo stars, astronomers can infer the conditions present in the early stages of the Milky Way's formation (Beers T. C 2014).

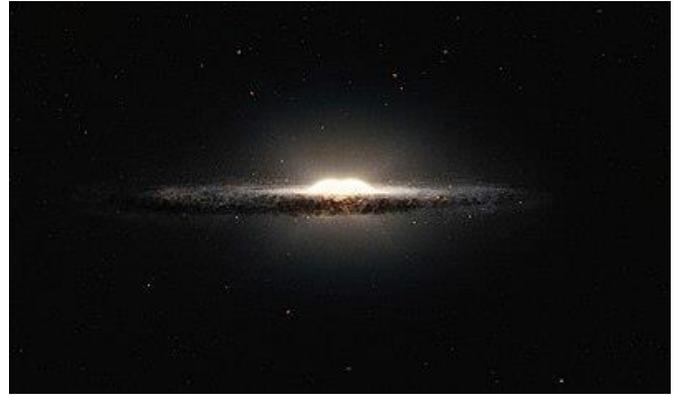

**Fig 2**. Artist's impression of the central bulge of the Milky Way.

**Bulge Stars:** The central bulge of the Milky Way contains a dense concentration of stars, forming a distinct structure within the galaxy. Bulge stars are characterized by their older ages and relatively high metallicities. The bulge likely formed through processes involving rapid star formation, possibly due to gas inflows and interactions. Studying bulge stars offers insights into the central regions of the galaxy, including its dynamical properties, star formation history, and potential connections to the growth of supermassive black holes (Genzel & Eisenhauer 2010).

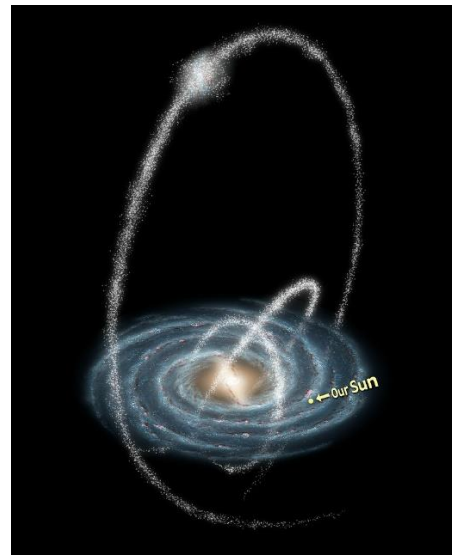

**Fig 3**. Stellar streams in the Milky Way, discovered in 2007.

**Stellar Streams and Substructures:** Stellar streams and substructures are intricate patterns of stars that result from the disruption of smaller satellite galaxies or star clusters. These structures can be found throughout the Milky Way and provide valuable information about past accretion events and interactions. By tracing the orbits and properties of stars within these streams, astronomers can infer the properties of the progenitor systems and the mechanisms responsible for their disruption. Stellar streams offer a window into the dynamic history of the Milky Way's interactions with its satellite galaxies (Johnston, K. V 2008).

# 2 SPECTROSCOPIC TECHNIQUES IN GALACTIC ARCHAEOLOGY

Spectroscopic techniques play a pivotal role in modern astronomy, enabling scientists to unravel the complex nature of celestial objects by analyzing their light spectra. In the context of galactic archaeology, these techniques have revolutionized our understanding of the Milky Way's formation and evolution. This report explores three crucial spectroscopic techniques used in galactic archaeology: radial velocity measurements, chemical abundance analysis, and stellar age determination (Hawkins, K 2016).

**Radial Velocity Measurements:** Radial velocity is a fundamental parameter that characterizes the motion of an object along the line of sight. In galactic archaeology, radial velocity measurements provide insights into the kinematics and dynamics of stars within the Milky Way. These measurements are achieved by analyzing the Doppler shift of spectral lines in a star's spectrum. As stars move toward or away from an observer, their spectral lines experience a shift towards the blue (blue-shift) or red (red-shift) end of the spectrum, respectively.

Radial velocity measurements are crucial for identifying different galactic components such as thin and thick disk stars, halo stars, and stars in stellar streams. By mapping the velocity distribution of these components, astronomers can infer the overall motion of stars in the Milky Way, shedding light on its assembly history, gravitational interactions, and potential merger events (Koposov, S. E., 2019).

**Chemical Abundance Analysis:** Chemical abundance analysis involves measuring the relative abundances of different elements in a star's atmosphere. The composition of elements in a star's spectrum provides valuable information about its formation environment and evolutionary history. In galactic archaeology, chemical abundance analysis offers insights into the nucleosynthesis processes that occurred during a star's birth.

By studying the ratios of various elements, astronomers can distinguish between stars from different galactic components and infer their birthplaces. For instance, halo stars typically exhibit lower metallicity compared to disk stars, indicating distinct formation scenarios. This technique also helps trace the enrichment history of the interstellar medium and offers clues about the chemical evolution of the Milky Way over time (Nissen & Schuster 2010).

**Stellar Age Determination:** Stellar age determination is a challenging yet crucial aspect of galactic archaeology. The age of a star is a key parameter for understanding its role in the galaxy's evolutionary history. Spectroscopic techniques are employed to estimate stellar ages through methods such as isochrone fitting and asteroseismology. These techniques involve comparing observed spectral features to theoretical models and evolutionary tracks.

Stellar age determination provides insights into the chronology of different galactic components and the timescales of their formation. It aids in distinguishing between stars that belong to the Milky Way's various generations and offers a way to explore the temporal evolution of galactic structure (Feltzing, & Chiba 2013).

# 3 STELLAR KINEMATICS AND DYNAMICS IN GALACTIC ARCHAEOLOGY

Stellar kinematics and dynamics provide a window into the complex motion of stars within the Milky Way galaxy. These aspects play a pivotal role in galactic archaeology by offering insights into the history of star formation, the evolution of galactic structures, and the interactions that have shaped our galaxy over cosmic timescales. This report explores three fundamental aspects of stellar kinematics and dynamics in the context of galactic archaeology: orbits and trajectories of stars, galactic rotation and circular velocity, and velocity dispersion profiles (Allende Prieto 2006).

**Orbits and Trajectories of Stars:** The motion of stars within the Milky Way is governed by the gravitational potential of the galaxy. Understanding the orbits and trajectories of stars is crucial for deciphering their histories. By tracing the paths of stars over time, astronomers can infer their origins, interactions, and potential future destinations. The study of stellar orbits allows for the identification of stars that may have been accreted from smaller satellite galaxies or have been perturbed by gravitational interactions.

Advanced techniques, such as N-body simulations and dynamical modeling, enable astronomers to reconstruct the past and predict the future motions of stars. These methods provide insights into the assembly history of the Milky Way, including mergers, tidal interactions, and dynamical processes that have led to the formation of distinct galactic components (Sofue Y 2012).

**Galactic Rotation and Circular Velocity:** The Milky Way exhibits rotation, wherein stars orbit the galactic center. Galactic rotation is a fundamental parameter that underpins the kinematic structure of the Milky Way. By studying the velocities of stars in different parts of the galaxy, astronomers can map the rotation curve and determine the circular velocity profile. The circular velocity characterizes the balance between the gravitational pull from the galaxy's mass and the centripetal force due to rotation (McMillan 2011).

Understanding the circular velocity profile provides insights into the distribution of mass within the galaxy. Deviations from the expected rotation curve can indicate the presence of unseen dark matter and guide theories about its distribution. Galactic rotation also plays a crucial role in tracing the mass distribution of the Milky Way and inferring the properties of its central supermassive black hole.

**Velocity Dispersion Profiles:** Velocity dispersion refers to the spread of velocities among stars within a particular region of the galaxy. Different galactic components, such as the thin and thick disk, halo, and bulge, exhibit distinct velocity dispersion profiles. These profiles provide information about the dynamical state of each component and the processes that have influenced their formation and evolution.

Analyzing velocity dispersion profiles helps astronomers understand the dynamical equilibrium of the galaxy, the effects of resonances, and the interactions between different components. It also provides insights into the heating mechanisms that lead to the dispersion of stars' velocities over time (Loebman 2018).

# 4 STELLAR AGE DISTRIBUTION IN GALACTIC ARCHAEOLOGY

The age distribution of stars within the Milky Way holds the key to unraveling its formation history and evolutionary processes. Stellar ages provide insights into the chronology of star formation events, the interplay of galactic components, and the underlying mechanisms that have shaped our galaxy over cosmic timescales. This report delves into three crucial aspects of the stellar age distribution in the context of galactic archaeology: the stellar age-metallicity relation, age determination methods (isochrone fitting and asteroseismology), and the chronology of different galactic components (Minchev & Famaey 2010).

**Stellar Age-Metallicity Relation:**

The stellar age-metallicity relation is a cornerstone in galactic archaeology, offering a link between the chemical evolution of the Milky Way and its star formation history. This relation describes how the metallicity of stars (i.e., the abundance of elements heavier than hydrogen and helium) evolves with their age. Studying this relation provides insights into the enrichment history of the interstellar medium, the timescales of star formation, and the role of various processes in shaping the galaxy (Feltzing & Gilmore 2000).

**Age Determination Methods:**

Determining the ages of stars is a complex yet vital task. Two prominent methods for stellar age determination are isochrone fitting and asteroseismology. Isochrone fitting involves comparing the observed properties of stars (e.g., brightness, color, metallicity) to theoretical isochrones representing different ages and metallicities. Asteroseismology leverages the study of stellar oscillations to infer internal properties, leading to precise age estimates (Marigo 2017).

**Chronology of Different Galactic Components:**

Stellar age distribution aids in reconstructing the chronology of different galactic components. Thin and thick disk stars, halo stars, and bulge stars have distinct age profiles that reflect their formation epochs and dynamics. Younger stars predominantly populate the thin disk, while the thick disk and halo host older stars. This temporal differentiation is crucial for understanding the assembly of the Milky Way and the role of various galactic processes, such as mergers, accretion, and dynamical interactions (Silva Aguirre 2015).

# 5 RESULTS AND DISCUSSION

In the realm of galactic archaeology, the classification of stellar populations has revealed profound insights into the composition, origin, and dynamics of the Milky Way. The categorization of stars into thin and thick disk, halo, and bulge components offers a framework to understand the diverse histories of these galactic constituents. The distribution and kinematics of these populations reflect distinct formation scenarios, with thin disk stars representing a younger, dynamically cooler population characterized by higher metallicity, and thick disk and halo stars comprising an older, dynamically hotter population with lower metallicity. This classification informs us about the assembly processes, mergers, and interactions that have shaped the Milky Way's evolution over time.

Spectroscopic techniques have emerged as indispensable tools in deciphering the intricate details of stellar properties, abundances, and dynamics. Radial velocity measurements enable the determination of star velocities along the line of sight, unveiling the overall kinematic structure of the galaxy. Chemical abundance analysis, through the study of element ratios in stellar spectra, provides insights into nucleosynthesis processes and the history of star formation. Stellar age determination methods, including isochrone fitting and asteroseismology, allow us to unravel the ages of stars and reconstruct the temporal chronology of the galaxy. These techniques, coupled with advances in observational technologies, are crucial for characterizing stellar populations and enhancing our understanding of the Milky Way's formation.

Stellar kinematics and dynamics offer a profound glimpse into the motion and gravitational interactions of stars within the Milky Way. The study of orbits, trajectories, and velocity dispersion profiles contributes to reconstructing the assembly history of the galaxy. Galactic rotation and the circular velocity curve provide essential information about the mass distribution, dark matter content, and potential presence of central supermassive black holes. The velocity dispersion profiles of different galactic components unveil the dynamical equilibrium, resonances, and heating mechanisms operating within the galaxy. These insights enable us to discern the interplay between different structural components and their interactions over cosmic timescales.

The age distribution of stars is a linchpin for understanding the chronology of star formation events and the evolution of various galactic components. The age-metallicity relation, established through stellar ages and metallicity measurements, provides insights into the interplay between chemical enrichment and star formation history. Isochrone fitting and asteroseismology techniques enable us to estimate stellar ages with precision, facilitating the classification of stars into different generations. This chronology aids in deciphering the roles of thin and thick disk stars, halo stars, and bulge stars in the Milky Way's assembly. It also sheds light on the processes of mergers, accretion, and dynamical interactions that have contributed to the galaxy's complexity.

In conclusion, galactic archaeology, as a multidisciplinary endeavor, has unraveled a wealth of knowledge about the Milky Way's history and evolution through the analysis of stellar populations. The results presented in this report highlight the crucial role of stellar classification, spectroscopic techniques, kinematics, and age distribution in shedding light on the intricate narrative of our cosmic home. These insights contribute to a deeper understanding of the broader mechanisms governing galaxy formation and evolution in the universe.

# 6 CONCLUSION

Galactic archaeology, a complex endeavor involving the study of stellar populations, has illuminated the intricate history and evolution of the Milky Way. Synthesizing knowledge from diverse fields such as astronomy, physics, and chemistry, it has led to a profound understanding of the galaxy's composition, dynamics, and origins. Stellar population classification, distinguishing thin and thick disk stars, halo stars, and bulge stars, each with unique characteristics that narrate their histories, forms the foundation for understanding the galaxy's structure. Spectroscopic techniques, enabling precise analysis of star properties and compositions, have revolutionized our comprehension, with radial velocity measurements, chemical abundance analyses, and stellar age determinations unraveling the mysteries within stellar spectra and enhancing our cosmic narrative. Studying stellar kinematics and dynamics has unveiled the galaxy's assembly history, dark matter distribution, and the presence of a central supermassive black hole, revealing a dynamic interplay of forces that shaped the Milky Way's present state. The distribution of stellar ages, complemented by the age-metallicity relation and methods like isochrone fitting and asteroseismology, has unveiled the roles of distinct galactic components and intricate interplay of processes. In conclusion, galactic archaeology merges disciplines, weaving a tapestry that chronicles the Milky Way's evolution. Classification, spectroscopy, kinematics, and chronology interweave, unraveling mysteries, deepening our cosmic appreciation, and propelling our pursuit to understand the universe's farthest reaches.